# Title

# Parallel Seismic Data Processing Performance with Cloud-based Storage

## Full Author List

Sasmita Mohapatra[3], Weiming Yang[1], Zhengtang Yang[1], Chenxiao Wang[1], Jinxin Ma[2], Gary L. Pavlis[4], Yinzhi Wang[1]

1. Texas Advanced Computing Center, UT- Austin

2. UT - Austin

3. High Performance Research Computing, ORI, UT-Dallas

4. Department of Earth and Atmospheric Sciences, Indiana University, Bloomington, IN 47405

## Corresponding author indicated and contact information provided

Sasmita Mohapatra

Sasmita.mohapatra5@gmail.com

sasmita@utdallas.edu

## Conflict of interest statement

The authors acknowledge there are no conflicts of interest recorded.

## Abstract

This article introduces a general processing framework to effectively utilize waveform data stored on modern cloud platforms. The focus is hybrid processing schemes where a local system drives processing. We show that downloading files and doing all processing locally is problematic even when the local system is a high-performance compute (HPC) cluster. Benchmark tests with parallel processing show that approach always creates a bottleneck as the volume of data being handled increases with more processes pulling data. We find a hybrid model where processing to reduce the volume of data transferred from the cloud servers to the local system can dramatically improve processing time. Tests implemented with Massively Parallel Analysis System for Seismology (MsPASS) utilizing Amazon Web Service's Lambda service yield throughput comparable to processing day files on a local HPC file system. Given the ongoing migration of seismology data to cloud storage, our results show doing some or all processing on the cloud will be essential for any processing involving large volumes of data.

## Introduction

Cloud computing is revolutionizing scientific computing with large datasets. Information Technology jargon and blatant sales pitches found on the internet obscure some simple facts about what "cloud computing" really means. Some anchor points that may prove useful to this community are:

1. A cloud system is conceptually the same as high-performance computing (HPC) clusters operated by many universities and national centers like the Texas Advanced Computing Center (TACC). Both are collections (clusters) of many interconnected computers. The main difference is that the hardware and software of cloud systems is optimized for data handling while HPC systems are tuned to running large simulation jobs.

2. The software that manages scheduling jobs on a cloud system and an HPC system are radically different. That means not only CPU and memory resources but how data is accessed by both cloud and HPC systems.

3. A fundamental concept in all current generation cloud systems is "virtualization". A simple perspective on what that means is that the definition of all the components that define the cluster are abstracted as "virtual machines" that define the environment a workflow runs in. In contrast, virtual machines can be used in an HPC environment, but they are not essential.

The key point is that a cloud system is a platform for massively parallel computing, but a cloud system has a different software and hardware environments than HPC clusters.

Until recently the software infrastructure of seismology had no generic capability of handling large data sets with parallel processing at all, let alone modern cluster systems. Existing software like ObsPy and SAC are designed to process one waveform file at a time, which becomes inefficient for large datasets. This file-by-file approach limits scalability and slows down seismic analysis workflows. The rise of cloud computing also fueled the development of data formats optimal for cloud native storage and parallel throughput. The combination of cloud-native data formats and horizontal scaling of cloud computing architecture is an attractive solution for seismological research. It is impacted by the recent emergence of applications of big data, cloud computing, and machine learning to earthquake seismology. Addair et al. (2014) first demonstrated the applicability of the Hadoop software framework to large-scale seismic data processing. Articles by Mohammadzaheri et al., 2013; Dodge and Walter, 2015; Chen et al., 2016; Magana-Zook et al., 2016; Junek et al., 2017; Choubik et al., 2020; Clements and Denolle, 2020 demonstrate that the MapReduce concept, is an efficient programming model for many forms of seismic data processing. A large fraction of seismology data processing amounts to application of a sequence of algorithms to a subsets of data. Such processing workflows are well suited for parallelization, in general, and the optimal design for cloud computing, in particular, as discussed by MacCarthy et al., 2020.

The volume of data available for seismology research has outgrown the software infrastructure for handling that data. One of the major bottlenecks today is data delivery from the current generation of archival storage. MacCarthy et al. (2020), all report that downloading all three-component, broadband seismic data from the USArray Transportable Array (TA) would take nearly a month to download, and "US Regional" data (from IRIS DMC Data Statistics, see Data and Resources) would take a year, assuming uninterrupted acquisition. Our experience was even worse. Around the same time period we used ObsPy's mass download procedure (procedure (https://docs.obspy.org/packages/autogen/obspy.clients.fdsn.mass_downloader.html) to download all broadband data for teleseismic events within the footprint of the USArray deployment. It took us six months to complete that task. We concur with the assertion by MacCarthy et al. (2020), that these retrieval cost inhibits "survey scale" research and that the problem will continue to grow as high sample-rate instruments, such as nodal seismometers, and large research datasets become more common (Incorporated Research Institutions for Seismology, 2019). MacCarthy et al. (2020) also argue that one solution to this problem is handling raw data at its point of storage using cloud computing. This paper explores that issue by considering tradeoffs in performance with respect to the kinds of processing best done within a cloud system. We cast the problem in terms of generic solutions for working with a system where data centers are hosted in the cloud. The generic models we describe here are an important contribution because we show how they can be made concrete using the MsPASS software framework (Wang et. al, 2021). In addition, the performance tests we describe using MsPASS and Amazon Web Service (AWS) provide important baseline data for designing data processing workflows that can effectively utilize cloud storage.

Although EarthScope is in the process of migrating all the seismic archive to the cloud, the testing we report here was done on the only large data set currently openly available on the cloud at the time this work was completed; the Southern California Earthquake Data Center (SCEDC) archive hosted on AWS (Yu et al, 2021). In addition to EarthScope, other U.S. data centers like the Northern California Earthquake Data Center (NCEDC) are also contributing to this shift, with resources such as their AWS Open Dataset for Northern California Earthquakes. We find that in that environment there is at least an order of magnitude increase in throughput if the data are windowed using the AWS "lambda" capability before the data are transferred to local storage. The implications are not surprising but important. That is, the unambiguous role of cloud computing, at present, is preprocessing the data as much as possible to reduce the volume of data to be transferred over a long-haul internet connection to a home institution.

## Conceptual Models for Modern Data Processing

This paper addresses a small subset of the larger world of cloud computing that is current interest to most seismologists. The generic issue is how the community can make a more effective use of the large archive of seismology data managed by EarthScope and other groups now being moved to the cloud. The results of this paper apply to that problem with these assumptions to reduce the scope of the problem:

1. The size of the waveform archive is the primary issue. Everything else EarthScope curates is tiny by comparison.
2. A critical community need is multiple ways to access that data to support research. Mission agencies with fixed data requirements are secondary. i.e. in this paper we are addressing issues important for a scientist working on basic research problems not the needs of an operational system like a seismic network operator.
3. All scientists have home computing systems where they need to do their research. Research needs on that system can span a range of tasks that could be done on a desktop or laptop computer to something that would require the most advanced HPC cluster. The key assumption is "home" today is almost guaranteed to not be on the cloud. Furthermore, there will always be a role of a local system to produce the final products that go into a published paper, so all processing eventually ends up with at least one stop occurring on premises.

4. The reason data centers like EarthScope and assorted FDSN (Federation of Digital Seismograph Networks) data centers exist is to support science using the data they curate. The data holdings of all such centers is huge compared to what most, if not all, institutions could absorb. For that reason a universal first step in any research project is selecting a subset of data from the archive. That starts as a recipe for what needs to be selected. It continues, however, in stages until a product is produced that forms the basis for a scientific result.

With those assumptions, we suggest that every example we know of for interaction with data centers can be reduced to three distinct tasks: (1) read some data from the archive, (2) do some first-order processing of the waveform data, and (3) save the data to the "home" system for the research project of interest. There are two end members implementing this conceptual model. The first, which is the status quo today, can be called the "download model". That is, step 1 is a request for data from the archive that is delivered by some mechanism to local storage. One then does all waveform processing on their "home" system, so the entire waveform subset needs too at least reside temporarily on the home system. The other end member is to just do all your work on the cloud. Although that end member is feasible today, there is little question the training barrier required for most of the community is currently far too high for that to be a realistic solution. That may change with facilities like the GeoLab gateway at EarthScope and related cloud training material (https://arxiv.org/abs/2409.19147). At this time, however, that barrier is real. In addition, as we discuss later there are major economic implications for that model as all cloud systems are commercial entities. As a result, we assert that for the foreseeable future the optimal solution is a hybrid where "first-order processing" is between the end members of nothing for the download model and everything for the cloud only model.

With that background, what defines "first-order processing"? Some examples are predictable and EarthScope and other data centers may well provide turnkey solutions for them. Examples are prewindowed event data or some basic signal processing steps like demean operators. A fundamental issue, however, is that research by definition needs to be flexible, and turnkey solutions may limit innovation. As a result, the range of what defines "first-order" processing is problem dependent and thus needs to be flexible.

In this paper we describe tests relevant to the two conceptual models of hybrid cloud processing illustrated in Figure 1. Figure 1a and 1b illustrate what we will call the cloud storage model. This model is an instance of something most readers of this paper have likely experienced. That is, all cell phones and many laptop computers have tightly integrated cloud storage systems hidden behind applications that run on the device. Those applications do not store all the data they consume locally but fetch the fraction of data needed driven by a request. Future developments of data centers may provide other abstractions that provide alternatives but the model we used in this paper is the traditional file-system model. That is, we aim to access data stored in files on the AWS Simple Storage Service (S3) cloud storage. Other cloud systems have similar functionality, but with a different Application Programmer's Interface (API). The general idea is to use the cloud equivalent of low-level IO operations: (a) open file, (b) optional seek, (c) read, and (d) close file. That concept is illustrated in Figure 1a for a parallel processing instance with three "workers". In that context a scheduler, which is illustrated as box in that figure, sends instructions to each worker telling it which data file to read, how far to seek, and how many bytes it should read. A key point is this conceptual model uses an abstraction that allows interaction of the cloud store as if it were a local file system.

Figure 1b illustrates a second conceptual model for cloud processing we will refer to as the “shared processing model”. The basic idea is to offload some or all waveform processing to the cloud system. Figure 1b illustrates that idea by showing processing “workers” running on both the “home” and “cloud system”. In this model a particular waveform is handled as follows:

1. The scheduler sends instructions to a cloud-side worker to read the data that defines a particular waveform stored in the archive.
2. A reader running on the cloud system loads the requested data into its memory.
3. The cloud instance runs a sequence of processing algorithms on that waveform.
4. The result of the cloud processing is returned via an open internet connection to the local system. Note in this paper that it is assumed to be a seismic data object, but it wouldn’t have to be e.g. it could be an arrival time estimate produced by an automated picker.
5. The scheduler passes the return to a local worker process. That worker may do additional work on the result or just write it to local storage.

Note the approach in Figure 1b makes the most sense if step 3 produces a significant reduction in data volume that is returned through the internet connection to the “home system”.

## Prototype Implementations:

## Software Framework

A fundamental problem today with working on cloud systems is most of the software in common use in our community is not compatible with constraints imposed by the environment. Even worse is that fact that even if one made the effort to port a common tool like SAC to the cloud environment, it would be fundamentally incapable of fully exploiting the massive parallelization capabilities that characterize cloud systems. The only generic solution to this problem that exists today is the MsPASS framework (Wang et al., 2021) we used for the work reported in this paper. Features of MsPASS that made it the tool of choice for this study are:

1. MsPASS is containerized, which is the standard method today to create the virtual environment that defines a cloud system.
2. MsPASS parallelism is scalable and can run on a single desktop machine to a cluster like AWS with nearly unlimited numbers of processors.
3. The portability of the package makes it possible to run MsPASS on both ends of hybrid setups like that illustrated in Figure 1b.
4. MsPASS has an integrated database system that is essential for managing large data volumes.

## Data Index Abstraction

We used the MsPASS framework (Wang et al., 2021) to implement prototypes for the two cloud data processing models illustrated in Figure 1. The tests all accessed the SCEDC data set stored on AWS (MacCarthy (a) & (b) et al. (2020), Yu et. al. (2021)). The archive contains waveform data from the southern California Seismic Network going back to the year 2000. The data are stored as miniseed day files and event files with a total data volume ~100TB. The storage is abstracted as what Amazon calls an "S3 bucket" (Simple Storage Service).

The first step for interacting with data stored on S3 is to build an index in order to enable efficient querying and access to relevant waveform files. We treated this issue as a problem independently from data processing. The assumption is that if one were doing a full-scale data preprocessing workflow, the index would be already available by some mechanism. For our tests we had to build one, but the expectation is that a more efficient mechanism will need to be developed in the future to supply the index for the massive EarthScope and related FDSN archives.

For the SCEDC data we developed a prototype for building an index to their continuous and event files stored on S3. The prototype code was implemented as part of the MsPASS Database class used in MsPASS to abstract all waveform IO operations. In particular, the index we constructed is stored as a MongoDB document that when loaded using the python API (PyMongo) creates a python dictionary that can be used to drive the reader. The Database class method called read_data abstracts the read operation to load a single channel of data stored in a particular S3 file for each document defined in the indexing documents. The prototype is not generic, but once the EarthScope system is fully released we expect to make the prototype more generic to allow reading any miniseed format archive stored on S3. In particular, the MongoDB document index document contains all the information needed to access a particular waveform and is easily extended to support any S3 indexing scheme.

The abstraction we used allows us to read data on S3 with a python script. The script needs only instantiate an instance of a Database object that holds the index, query MongoDB to define the subset of data desired, and then construct the data set with one channel of waveform data per document returned by the query. The only complication in reading from S3 is that the script must define the credentials needed to access the requested data via a pair of "access keys" that AWS uses for access control. The tests described in this paper all use an index stored in a local (home) database. We note, however, that the abstraction we use could allow the query mechanism to be done by the data provider with a minor modification. That is, the provider need only allows read access to their waveform relational database. There are standard "big data" tools for querying relational database servers in the Dask (https://docs.dask.org/en/stable/generated/dask.dataframe.read_sql.html) and Spark (https://spark.apache.org/docs/3.5.3/sql-ref.html) schedulers used in MsPASS. Converting the output of an SQL server to input to drive a MsPASS workflow requires only a few lines of python code.

## Cloud Storage Prototype

For the SCEDC data we used Amazon's boto3 python package (https://boto3.amazonaws.com/v1/documentation/api/latest/index.html) to interact with AWS. The approach we used was more-or-less a cloud version of the download model. That is, the index document contains the equivalent of a file name that the reader uses to download the entire file it references into the local system's memory. The usage is very similar to the web service download functions in ObsPy that may be familiar to many readers. The MsPASS reader then translates the miniseed binary data into the internal MsPASS data structure called a "TimeSeries" object where additional processing functions can be applied to the data before it is saved locally. A serial version of that process is conceptually identical to an ObsPy processing loop driven by web service calls to download and process one waveform at a time. A key difference MsPASS provides is the parallel processing capability illustrated conceptually in Figure 1a. For the tests in this paper, we used the parallel scheduling framework called Dask (https://www.dask.org/). Parallel processing in our test is driven by a parallel data structure Dask calls a "bag", which is conceptually a python list of things that doesn't necessarily fit in system memory. The "scheduler" illustrated in Figure 1a, in this context, more-of-less sends waveform index documents to each "worker" process. Each worker then uses boto3 to download the file the document defines, process the result, and save the (reduced size) result locally.

## Shared Processing Prototype

The conceptual model of Figure 1b could be implemented in a number of ways. Our prototype implementation is illustrated in Figure 2. The prototype uses a feature in AWS that Amazon refers to as a "lambda service" (https://docs.aws.amazon.com/lambda/). Readers familiar with "lambda functions", which is standardized in most modern computing languages including python, can appreciate the name. Conceptually, AWS lambda allows a single python function to be callable as a remote service from an application. That function can be anything provided it is stateless, meaning all relevant parameters describing its behavior must be sent to it. In the implementation for this paper, that function is a MsPASS function driven by a waveform index document that defines an S3 object (file). For this paper that lambda function does three things: (1) use the input document to construct a MsPASS TimeSeries object from the miniseed data, (2) window the data from day files to a specified interval, and (3) return the windowed data to the local system. It is important to realize that step 2 was intentionally made simple so we could evaluate input-output performance. In most applications additional processing functions would likely be called within the lambda function. Windowing alone is best thought of as an end-member example. We discuss the implications of this below in terms of load balancing and economics after showing the results focused on input-output performance 1(b), 2.

The key difference between Figure 1b and Figure 2 is how the cloud system manages load balancing. The generic model of Figure 1b shows the general problem the overall system must handle. That is, in a parallel processing environment the workers on the local system need to be balanced with the workers on the cloud system Figure 1b and 2 shows the naïve concept of matching workers one-to-one on the local and cloud sides. The AWS lambda service, however, does a sensible thing. The AWS cloud management system takes care of the load balancing automatically. That is, it adds workers behind the scenes to run instances of the lambda function as requests come it. That scaling occurs automatically but is limited by economics. That is, the service scales up to a limit. Adding performance requires paying more money to Amazon. That idea is illustrated in Figure 2 by showing different numbers of workers on the two sides of the connection.

## Performance Tests

Our primary goal was (figure 3) to evaluate the throughput possible in a world with data stored on a cloud system like AWS. To address that we focus on a processing end member that does minimal processing but reduces data volume significantly. That is, our test workflow does one and only one data handling step: time windowing. Related benchmarks with MsPASS indicate that the time spent on the windowing function alone can be largely neglected. Specifically, we found the windowing function from a typical SCEDC day long waveform segment on TACC's Frontera system (Stanzione et al. (2020)) is ~ 67 microseconds (µs) per call.

The test results were designed to appraise performance scaling with the number of processors dedicated to processing. Hence, we scaled the workflow from a single worker to 384 workers distributed across up to eight compute nodes. For the TACC's Frontera system more than one node is used when the number of workers exceeds 56, which is the number of cores on a single node. We ran the tests using both the Dask and Spark scheduler to appraise their relative performance. The differences between Dask and Spark were, however, negligible and results here are reported only for the Dask runs.

Figures 3a shows elapsed time data on previously indexed day files run three different ways:

1. The figures tagged "S3" are results running in the mode of Figure 1a; the cloud storage prototype where the full day volume is downloaded for local processing.

2. The figures tagged "S3 lambda" use the shared processing prototype implemented with AWS lambda as illustrated in Figure 2.

3. The results labeled "Scratch" are a control test. That is, the "Scratch" timing data was produced by running the same windowing workflow but with the miniseed files previously loaded and indexed on TACC's Frontera cluster. The "Scratch" label is a detail that the file system on which the data were stored is the Lustre file system on Frontera with that label. Scratch is a large virtual disk farm shared by all Frontera users for staging large data sets. The point is that Frontera "Scratch" is a state-of-the-art local file system for an HPC cluster. The "Scratch" results thus provide a measure of ultimate performance we could expect the other tests to, at best, approach.

Our results illustrated in Figure 3a show elapsed time for windowing 5660-day files from the SCEDC open data stored on AWS S3 (MacCarthy et al. (2020), Yu et al. (2021)). The approximate total size of data accessed is 28 GB.

# Results

Figures 3a shows that the performance of all three workflows improve (elapsed time decreases) as the number of workers increases. The scaling in all cases, however, departs significantly from the ideal scaling line illustrated in the figure. That is not unexpected, but it is important for the reader to understand what we do and don't know about what shapes those curves.

In the context of Figure 3a, the elapsed time can be expressed as:

$$T = T_s(N) + \frac{T_p(N)}{N} \qquad (1)$$

where $T_s$ is a serial portion of the code that cannot be parallelized, $T_P$ is the time to execute a fully parallel algorithm, and *N* is the number of workers running the parallel algorithm. In our tests all are functions of *N*. In all cases $T_s$ is dominated by dask/spark scheduling overhead. Profiling we have done and the Dask documentation indicate the $T_s$ is probably under 1 s for all tests. If so, $T_s$ is likely small for all but the largest values of *N* for the control (scratch) test.

The factors controlling $T_p$ are different for the different tests and all, as noted, are largely controlled by IO performance. In all cases, however, $T_p$ can be expanded as:

$$T_p = T_{open} + T_{read} + T_{unpack} + T_w + T_{close} . \qquad (2)$$

where each of the terms in the sum on the right hand side defines a generic concept that is implemented in different ways in each of these tests. The term $T_{open}$ for a file system read (scratch) is the time required to "open" the file. For both S3 and S3 lambda we know of no way to separate $T_{open}$ from $T_{read}$, which is the time required to read the stream of bytes from storage into memory. $T_{unpack}$ is the time required to convert the compressed miniseed packets into an array in memory for processing and $T_w$ is the time to run the window function. Independent timing tests with 1000 trials show the median value of $T_{unpack}$ is 50 ms and the median value of $T_w$ , as noted earlier, is 0.05 ms. The sum $T_{unpack} + T_w$ is approximately 50 ms and we can assume it is nearly constant for all tests. Variations can be expected only for hardware differences, which as best can tell are not dramatically different between Frontera and the hardware at AWS where we ran these tests. Finally, $T_{close}$ is the time required to close a file or to end an AWS transaction. For the scratch control test $T_{close}$is the file close time which can be neglected. For S3 it can also be neglected as it is, at most, the time to release the memory held by the miniseed file image in memory. For S3 lambda $T_{close}$ is something very different. It is the time to transfer the output of the window function back to Frontera.

That framework provides a model to explain the form of the curves in Figure 3. Consider the control experiment first as that model is likely familiar to most readers. File open and close times on a system like Frontera are tiny and of the order of 1 ms or less in normal conditions. For these tests we used only the 40 sps day files which are of the order of 5 Mb in size with read times of the order of 10 ms. As noted above $T_{unpack} + T_w$ is approximately constant for all files. If all the terms were independent of *N* equation (1) says the elapsed time plot in Figure 3 would be approximately linear at low values of *N* but asymptotic to log $T_s$ as *N* gets large. For the "Scratch" control $T_s$ can be neglected and the ideal scaling on the log-log plot of Figure 3a would be a linear curve passing through the serial processing time (1 worker). The S3 curve begins to flatten at tens of workers, indicating diminishing returns in performance as parallelism increases and highlighting a departure from ideal linear scaling. The reason is best understood by the alternative representation of these data plotted in Figure 3b expressed as throughput = $D/T$ where $D$ is data volume (30 Gb for these tests) and $T$ is elapsed time. Even on an HPC cluster like Frontera 300+ processes actively doing IO operations can reduce throughput i.e. open, close, and read times all are likely to increase on an IO bound task like this as *N* increases. We conclude for the control experiments our tests with larger numbers of workers are saturating the Frontera scratch file systems read bandwidth.

The S3 and S3 lambda tests are subject to similar issues, but the controlling factors are different. For S3 we devised a set of independent tests to measure the different terms in equation (2). First, we ran independent measurement of $T_{open} + T_{read}$ without the complication of *N* workers screaming for data at the same time. We found such transfers took an average of 0.75 s with a variance of less than 1% in 1000 trials. Since $T_{unpack} + T_w$ is approximately constant (approximately 50 s) and $T_{close}$ is approximately zero, we can project an ideal scaling assuming $T_{unpack} + T_w$ is independent of *N* as the line illustrated in Figure 3a. The intercept is the S3 benchmark time with Dask running one worker. At $N = 1$ with $T_s$ and $T_p$ constant equation (1) reduces to $T_s + T_p$. We point out that simple relationship because to make the ideal scaling curve match S3 either $T_s$ or $T_p$ must be larger than our model values. The more important issue is that the S3 curve departs strongly from the model prediction which is close to linear on the log-log plot of Figure 3a. Like the control experiment, however, Figure 3b provides the most likely explanation for that scaling. That figure shows that the Scratch throughput flattens at approximately 300 Mb/s while the S3 throughput flattens at around 30 Mb/s. Both are of the order of the IO bandwidth through which those data are read: scratch file system and internet connection respectively on Frontera. Hence, we conclude that the performance for scratch and S3 in these tests are IO limited when *N* is of the order of 10 or more.

The limiting factor for the S3-lambda case is different. The justification for this claim is found in Figure 4. There we represent the timing data in terms of the data volume the readers on the two sides of the system illustrated in Figure 2 must handle in these tests. The "Frontera" curve shows the throughput that must be handled at Frontera compared to what AWS had to handle under the hood to feed that data to Frontera at that rate. The two curves have exactly the same form but are offset on the y-axis by a factor of 300/86400 (5 min/1 day). The volume of data transferred is drastically reduced because of the windowing operation applied to day volumes. Because the y-axis in Figure 2b is logarithmic the Frontera curve is then a simple downward shift from the AWS curve. A key point is that the Frontera data rate is several orders of magnitude less than the IO bandwidth of the internet bandwidth at Frontera as measured roughly by the S3 curve of Figure 3b. We plot the control data read rate for scratch in Figure 4 and note that curve is similar in shape and only slightly offset from larger than the curve with the tag "AWS". That suggests a similar saturation of the IO bandwidth is occurring in these tests on the AWS server side. Given Frontera's scratch is a state-of-the-art cluster file system, that hypothesis is reasonable.

An issue with the scaling of S3-lLambda that is important to understand at this point is the resources allocated to the lambda service at AWS. S3-Lambda automatically scales in response to incoming requests, adjusting to traffic volume while being billed based on execution time and memory usage. It has a 15-minute run limit, along with concurrency thresholds and potential cold starts that may cause latency. Cloud services like AWS are "elastic" meaning the performance can be improved by increasing resources dedicated to the task on AWS. Increased performance, however, comes at a cost. With this model, a research project using an AWS lambda service would need to pay more money if the project required improved performance to be feasible.

## Discussion and Conclusion

The first thing all readers should recognize is that the modified download model as described in Cloud storage prototype section (Figure 1a) "S3" is not an effective solution for processing very large data sets. A case in point is the extended USArray data set we mentioned in the introduction made up P wave records of all broadband stations in the lower 48 states for the USArray recording period. The data set is of the order of 40 million waveforms. It took us about 6 months to download that data set with Obspy and web services. From the timing data above if we did the same operation on Frontera with one worker using MsPASS and the S3 algorithm, the job would run for 462 days. Figure 3b shows that parallelization with the "S3" tests with a rate approximately 20 times faster than a single worker. Hence, that same data set would still take the order of 20 days to assemble with 50 active workers. That is feasible, but an important warning is the times could be orders of magnitude slower due to local network performance if run elsewhere. We ran these tests on Frontera which has about as fast a connection to the internet possible. Time to assemble a large data set like the extended USArray example would, at best, remain painfully long using the S3 approach. Hence, the version of hybrid models of Figure 1b or 2 seem essential for working with massive data sets.

The current fundamental problem to implement a hybrid processing scheme like Figure 1b or Figure 2 is software. Our tests show MsPASS can be used to do that with the AWS Lambda service. MsPASS is generic enough that other ways to implement the more general approach of Figure 1b are feasible. In any case, the implementation is currently far from easy as it requires understanding of all the pieces of a complex collection of software. One possible solution may be a science gateway under development by the SCOPED project (https://www.earthscope.org/gateway). Smaller data sets may be handled with EarthScope's GeoLab (https://www.earthscope.org/data/geolab/), but at present that gateway is limited to 4 processors and provides limited cloud store to save processed data. Until then those needing to handle huge data sets either need to be very patient or be hardy enough to work with the MsPASS framework.

Our tests were run on the SCEDC data, but how will this work with the unified archive now staged to AWS by EarthScope. The current development plans to use S3 bucket storage with a web-service API to query the archive to retrieve index information (Chad Trabant, 2025, personal communication). That setup will allow use of a lambda service like that in our prototype for cloud-side preprocessing. The main differences are in how the data index is acquired and differences in credential validation for EarthScope and SCEDC. We suggest the lambda service model used here should be a standard way to handle preprocessing of large data volumes to assemble a research ready data set. That approach is not well matched to the GeoLab gateway, but is possible with MsPASS as our tests demonstrate.

A final point about the hybrid model using lambda service or S3 is economics. AWS and other cloud providers are commercial entities that charge real money for their services. Someone has to pay and there is little doubt the days of unlimited, free data access will disappear with the transition to cloud storage. The cheapest solution as described in cloud storage prototype section for the foreseeable future is likely to remain the "S3" model of Figure 1a. In that case, the data center needs only feed data in respond to user requests and the costs are the fixed storage costs and any computing that the data center pays. The hybrid model of Figure 1b or 2 changes the economics. The cost for the AWS Lambda service, for example, scales with the maximum number of processors allocated when the service is launched. The greater the computation workload shifted to the cloud, the higher the cost the task will incur from the provider. Hence, the economics of a hybrid system will be a tradeoff of local fixed costs to maintain research computing hardware versus the cost of doing your work through a commercial provider. All you can predict for sure about that problem is it is subject to large changes with time. Transition to that model, however, seems inevitable as it is essential to provide a mechanism for processing large data volumes obtained from the international archives.

## Data and Resources

The documentation of MsPASS is available at https://www.mspass.org/user_manual/. The Massive Parallel Analysis System for Seismologists (MsPASS) source code is available at https://github.com/mspass-team/mspass, and the containerized distribution can be found in https://hub.docker.com/r/mspass/mspass. The information about GridFS is available at https://docs.mongodb.com/manual/core/gridfs. Comparison with Spark is available at https://docs.dask.org/en/latest/spark.html. Examples of using the MsPASS implementation for S3 on AWS can be found in https://www.mspass.org/user_manual/importing_data.html. Several examples of automated script of the lambda function can be found here: https://github.com/SCEDC/cloud/tree/master/pds-lambda-example. For basic usage of S3 bucket for SCEDC data can be found here: https://scedc.caltech.edu/data/getstarted-pds.html. For basic usage of lambda, one can refer to https://scedc.caltech.edu/data/lambda.html. The SCOPED Gateway enables users to run seismology applications on HPC/Cloud through tapis: https://seisscoped.org/tapis-ui.

## Acknowledgments

This work is supported by the National Science Foundation through award OAC-2103494 and OAC-1931352.

## Full mailing address for each co-author

Yinzhi Wang: iwang@tacc.utexas.edu,

Gary L. Pavlis: pavlis@indiana.edu,

Weiming Yang: weimyang@utexas.edu

Chenxiao Wang: chenxiaowang@utexas.edu

Zhengtang Yang: yangzhengtang@utexas.edu

Jinxin Ma: jinxin.ma@utexas.edu

## Complete Figure Captions list

Figure 1. Generic models for parallel data processing with cloud data storage prototype and shared processing prototype. (a) illustrates how parallel processing can work with data stored on the cloud and downloaded as needed. In the implementation described in the text this is the model we refer to as “S3” because the implementation uses AWS S3 cloud storage. The large box on the right illustrates the hardware of a system where processing is performed and the irregular polygon at the right is an abstraction a cloud system. The example illustrates 3 worker processes running on a local system that request data and retrieve files through a common internet connection to a cloud server. In this case, the parallel scheduler only needs to interact with the local processes as illustrated with the red lines with arrowheads. Workers then request data (reader box) from the cloud and they receive the data through a common internet connection as illustrated with the heavy black array. Processing functions (Processing box) may do some work on the data before saving the results to local storage (Writer box). (b) illustrates what we call a hybrid model where some of the processing is done on the cloud system before being transferred to a local system. The illustration is an example with three workers running on the local system and three workers running on the cloud system. That generic approach would allow different numbers of workers on both sides to optimize performance. A key difference in this model compared to (a) is that the scheduler running on the local system has to send control messages to manage not only local but remote processes running on the cloud system. That is illustrated here with dashed red lines with arrowheads on both the local and cloud side of the system connected via the internet connection. A difference in this model is the reader box is on the cloud side data would be reduced in size with the box labeled "Cloud System Processing" before being transmitted across the long distance internet connection.

Figure 2. Implementation of hybrid processing using AWS lambda service. This figure is directly comparable to the generic form of this concept illustrated in Figure 1b. Note the difference is that the cloud side is implemented through a specific approach available on AWS called a lambda service (see text). The lambda service automatically scales worker processes to handle load up to a limit imposed by a cost structure. We illustrate that here through two worker boxes labeled "W1" and "W2". Each AWS workers handle files assigned by at AWS by their schedule. The local side only sends requests illustrated here by an orange arrows labeled "Index document". In our implementation the index document drives a python reader running as the lambda function at AWS. In tests presented in the paper that function also windows data to reduce the size of what is returned. The output of the lambda function is returned to the local (TACC Frontera) side as illustrated by the black lines with arrowheads. As in Figure 1 the dashed, red lines with arrowheads illustrate control data paths used by the scheduler running on Frontera.

Figure 3. Performance results reading SCEDC day file data from AWS on Frontera with different processing approaches. All results are for the same processing sequence involving only reading, unpacking miniseed data, and windowing the data to a 5 minute long segment. The data files used for these tests are 40 sps day files, of the order of 5 MB size, with read times of the order of 10 ms, (a) Displays the data as elapsed time as a function of the number of serial and parallel workers. The tag name in the legend defines the three algorithms discussed in the text. The dashed red line is the S3 benchmark time with Dask running one worker. (b) is the same data as (a) but converted to throughput and plotted with a logarithmic y-axis and a linear x-axis. The ideal scaling curve is omitted. (b) has a second logarithmic axis on the right with the data converted to the number of files (objects) processed per second.

Figure 4. Processing data volume scaling for different algorithms. This plot is similar to Figure 3b, but the curves represent the data being handled by different instances of the processing code. As in Figure 3 "Scratch" is the data being read and processed on Frontera in the control experiment reading from a high-performance local file system. The curve labeled "Frontera" is the data being handled on Frontera when reading from AWS with the lambda service handling the miniseed unpacking and windowing. The curve labeled "AWS" is the data rate AWS has to sustain to feed the lambda data sent to the processes running on Frontera. The "Frontera" curve is the same as "AWS" but offset by a factor of the ratio 86400/300 (1 day/5 min).

# Figures

## Figure 1(a)

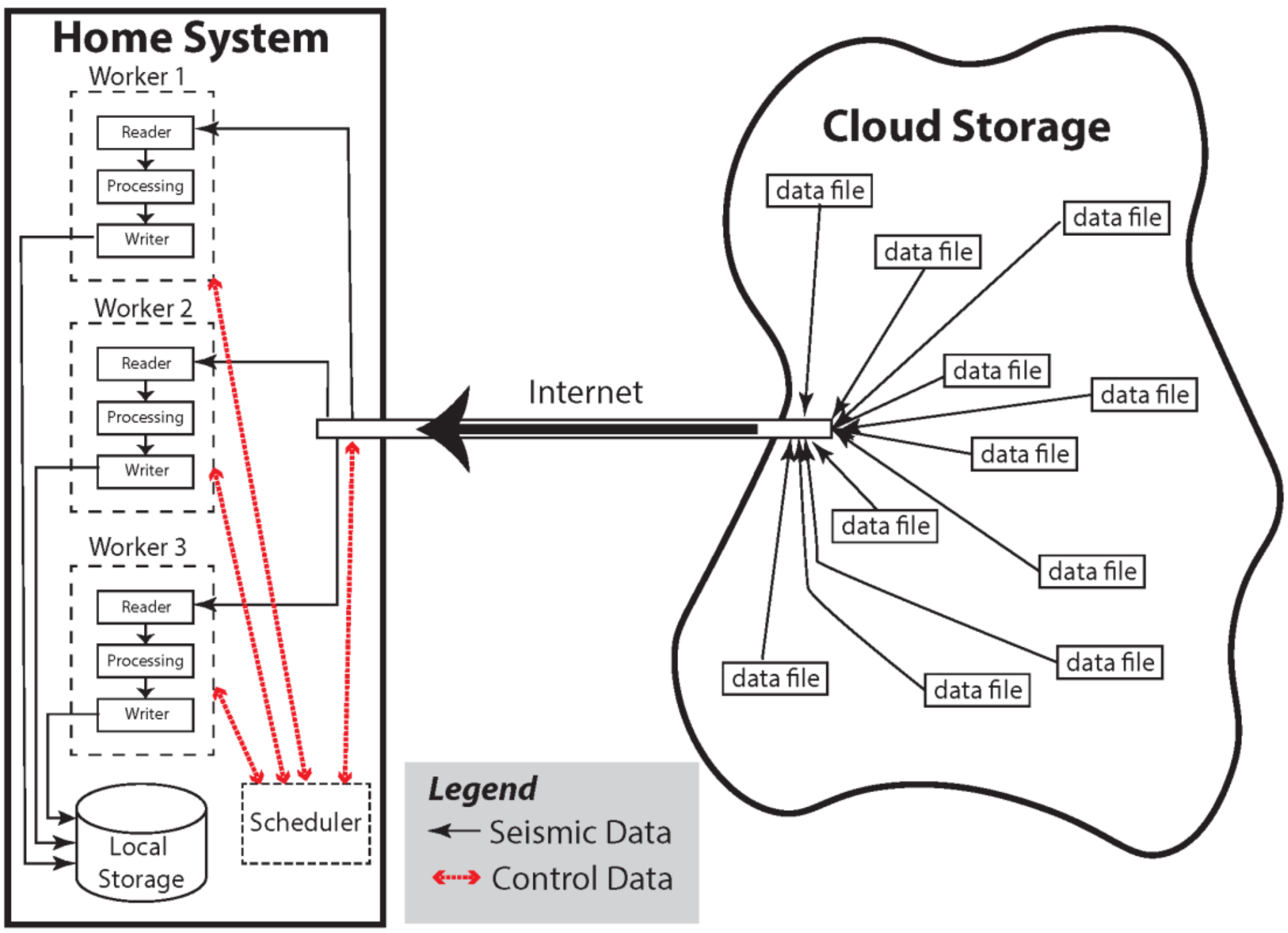

1(b)

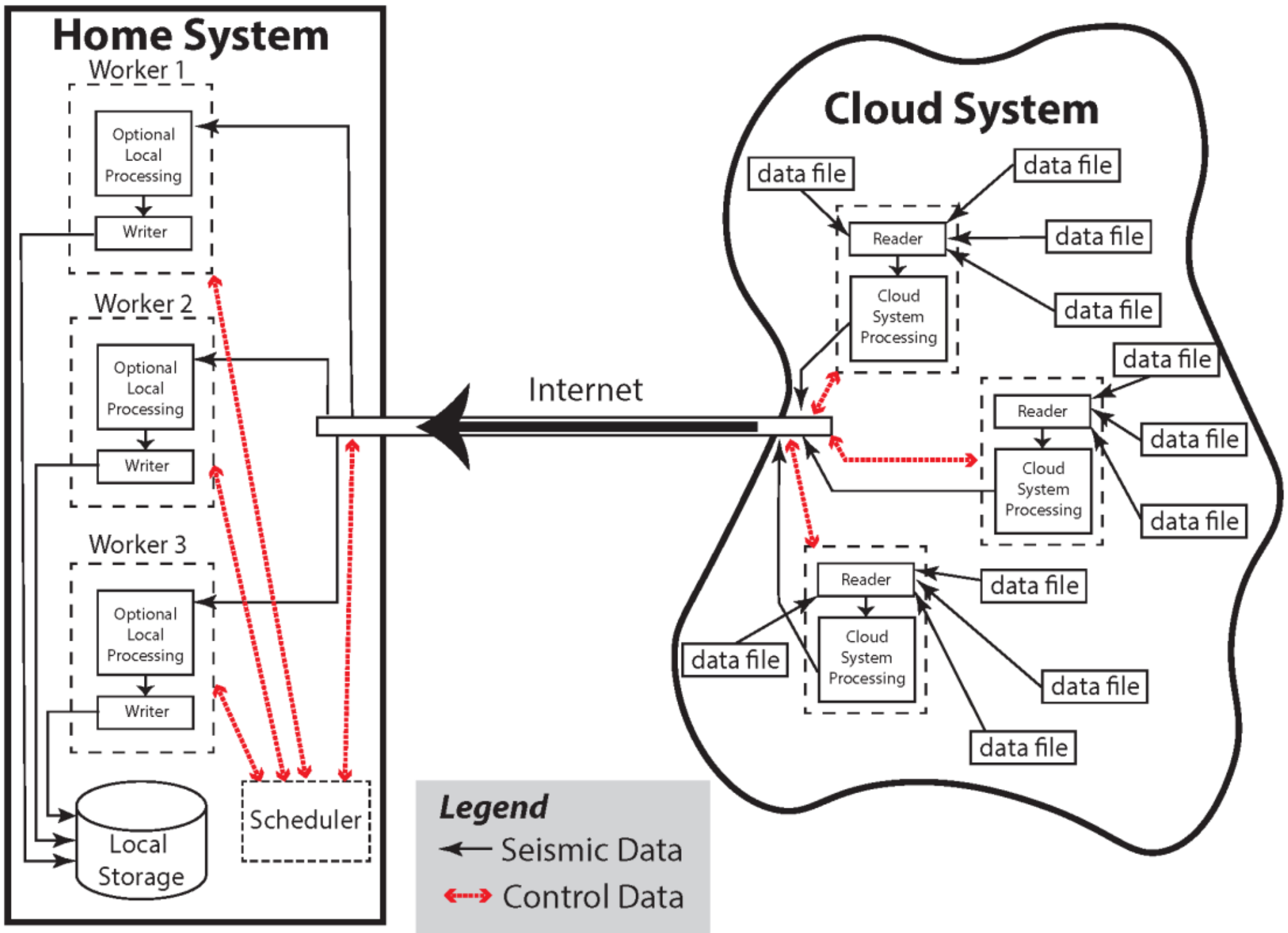
Home System
Worker 1
Optional Local Processing
Writer
Worker 2
Optional Local Processing
Writer
Worker 3
Optional Local Processing
Writer
Local Storage
Scheduler
Internet
Legend
Seismic Data
Control Data
Cloud System
data file
Reader
Cloud System Processing
data file
data file
data file
Reader
Cloud System Processing
data file
data file
data file
Reader
Cloud System Processing
data file
data file
data file
data file

Figure 2

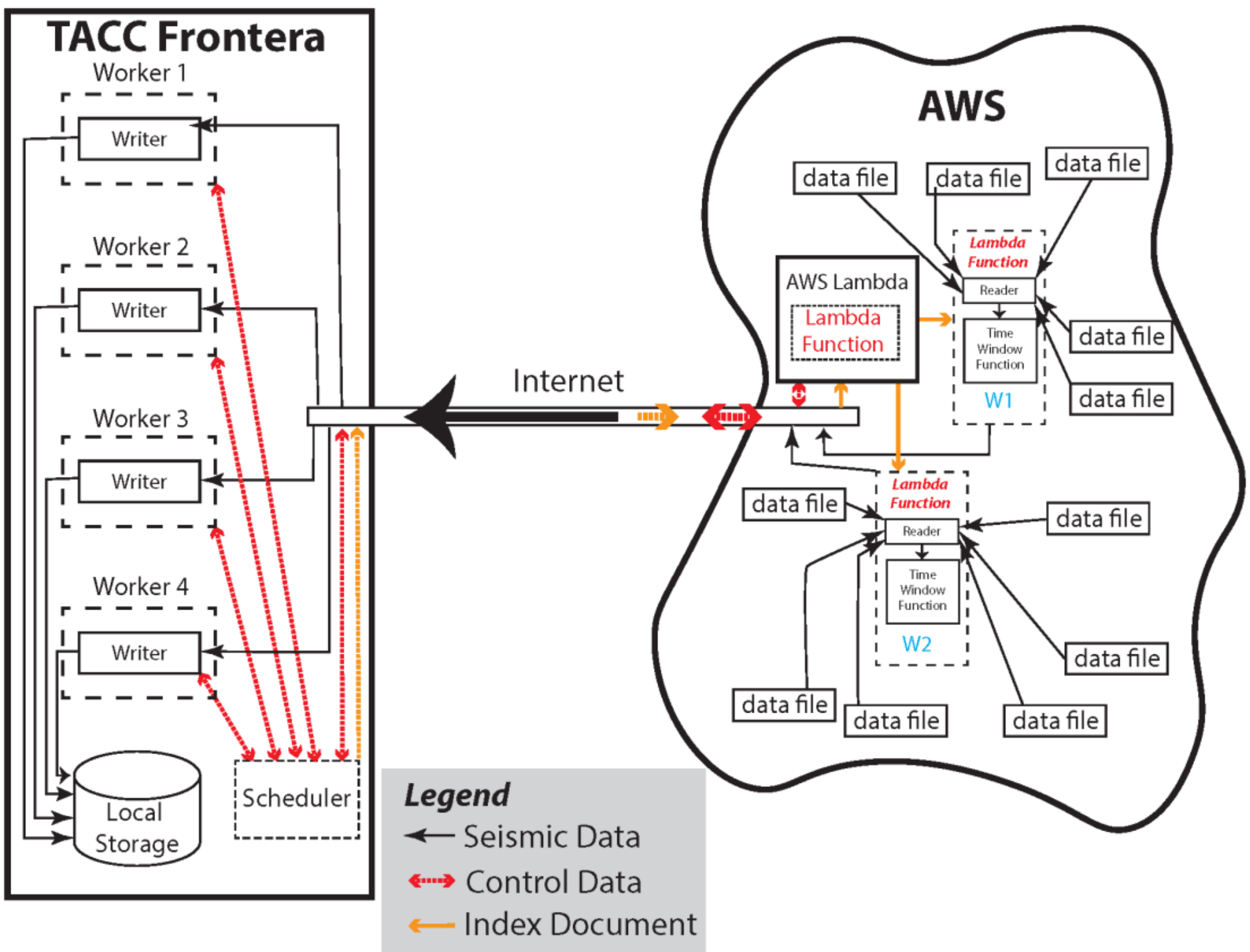
TACC Frontera
Worker 1
Writer
Worker 2
Writer
Worker 3
Writer
Worker 4
Writer
Local
Storage
Scheduler
Internet
AWS
data file
data file
data file
AWS Lambda
Lambda
Function
Lambda
Function
Reader
Time
Window
Function
W1
data file
data file
data file
Lambda
Function
Reader
Time
Window
Function
W2
data file
data file
data file
data file
data file
Legend
Seismic Data
Control Data
Index Document

Figure 3 (a)

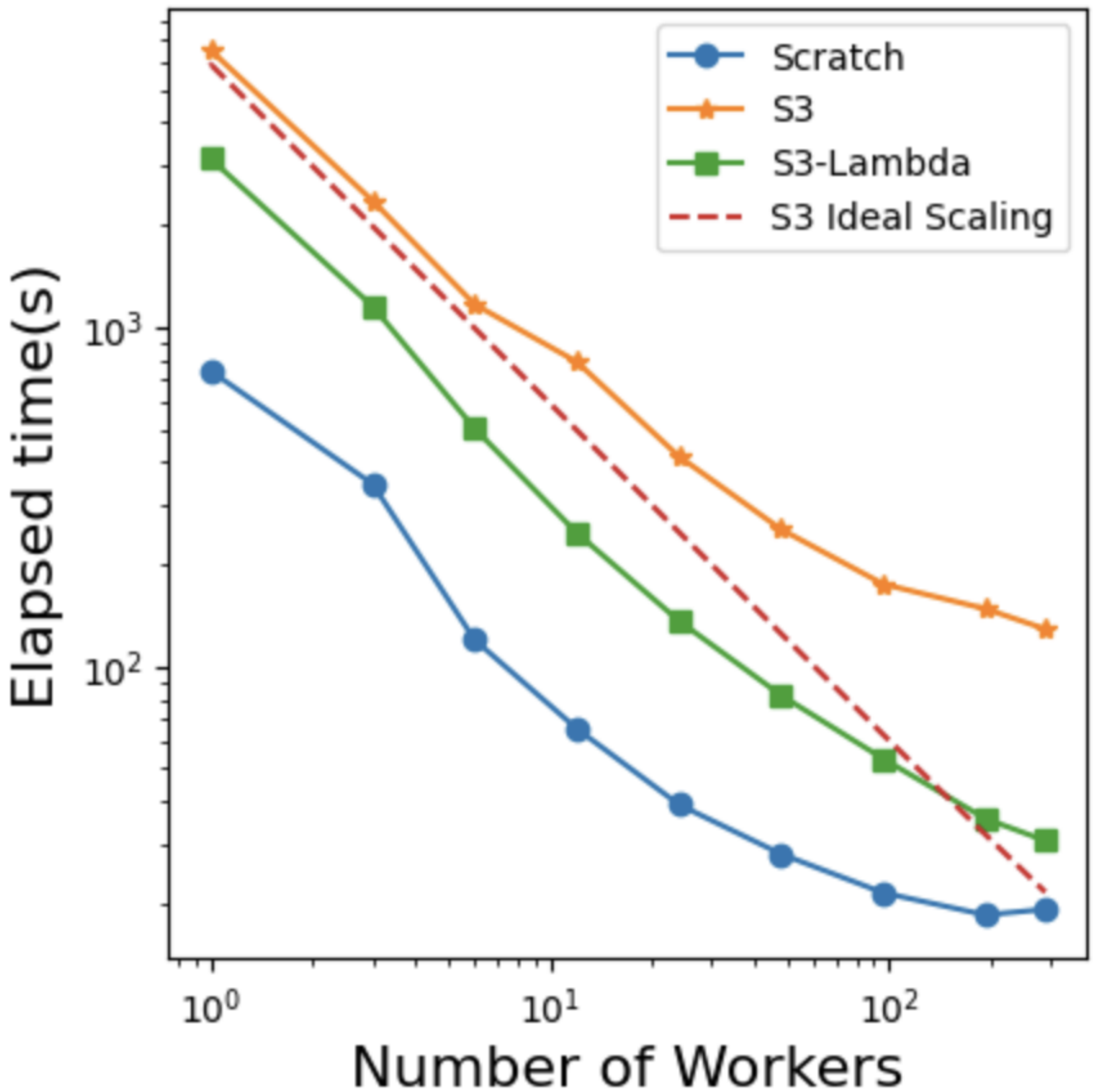
Scratch
S3
S3-Lambda
S3 Ideal Scaling
Elapsed time(s)
$10^3$
$10^2$
$10^0$
$10^1$
$10^2$
Number of Workers

Figure 3 (b)

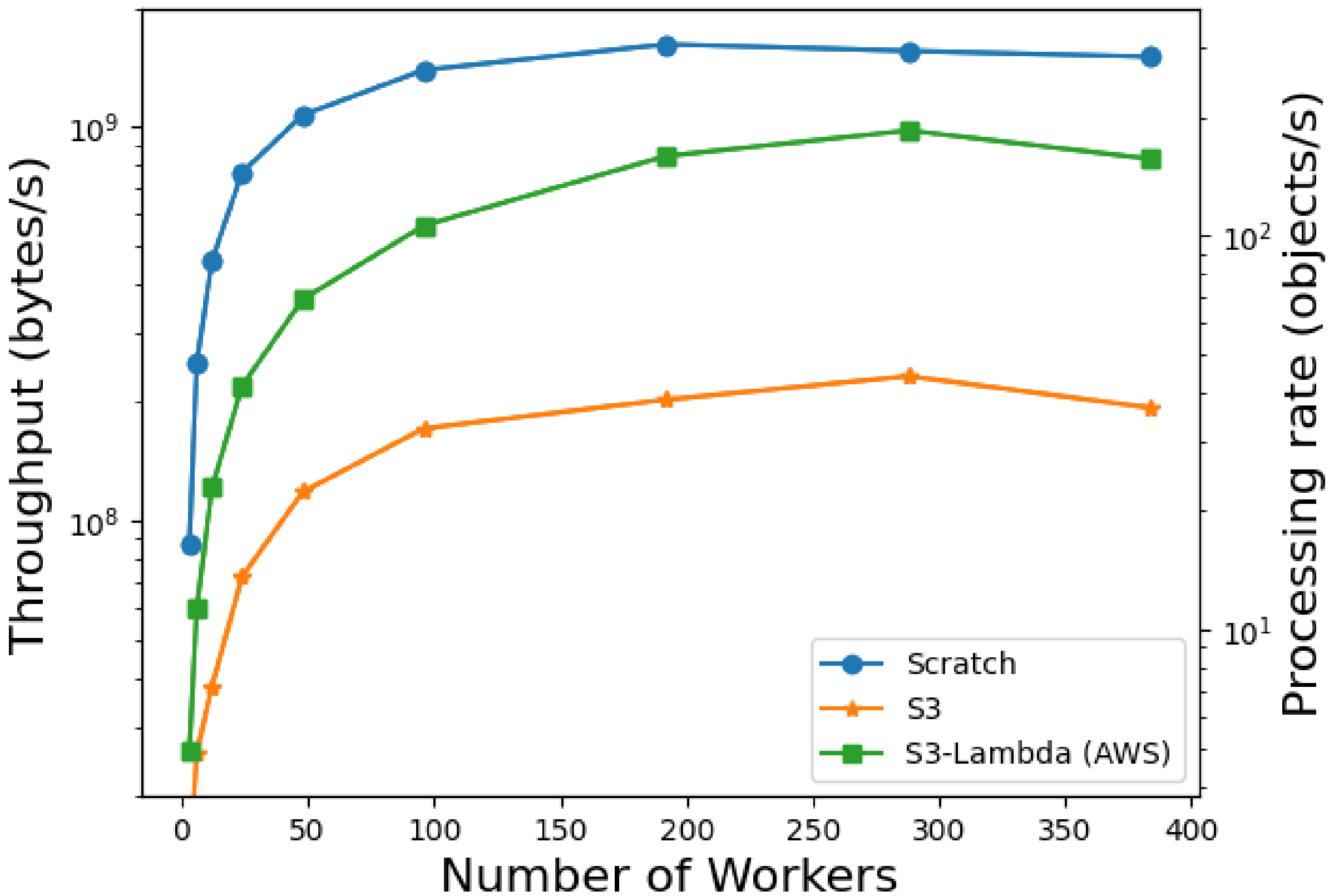

Throughput (bytes/s)
$10^9$
$10^8$
Processing rate (objects/s)
$10^2$
$10^1$
0
50
100
150
200
250
300
350
400
Number of Workers
Scratch
S3
S3-Lambda (AWS)

Figure 4

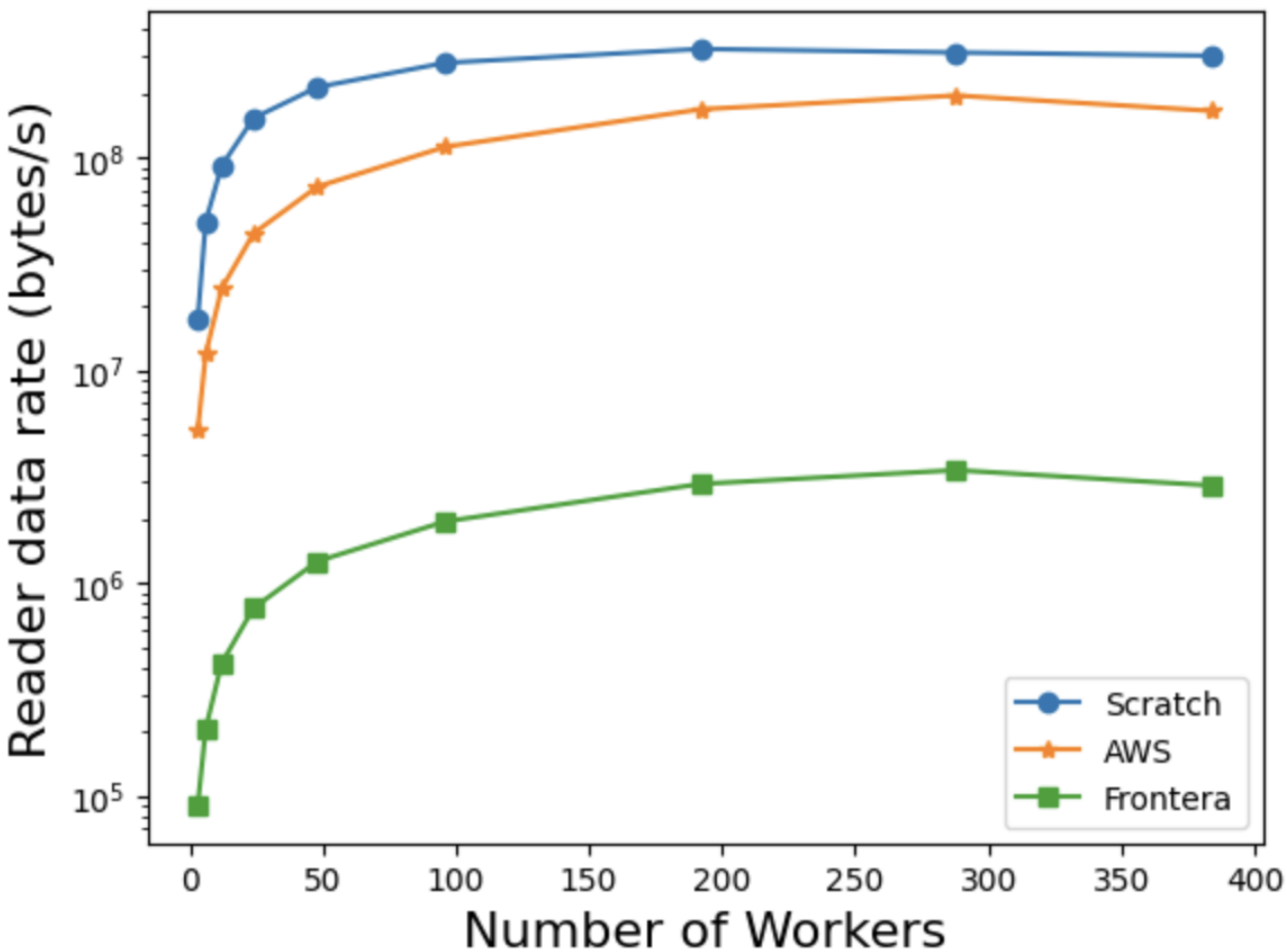
Reader data rate (bytes/s)
Number of Workers
Scratch
AWS
Frontera
0
50
100
150
200
250
300
350
400